# A DFT study on the mechanical properties of hydrogenated and fluorinated germanene sheets

M. Goli [1], S.M. Mozvashi [2], P. Aghdasi [1], Sh. Yousefi [1], R. Ansari[1,*]

[1] *Faculty of Mechanical Engineering, University of Guilan, P.O. Box 3756, Rasht, Iran*
[2] *Computational Nanophysics Laboratory (CNL), Department of Physics, University of Guilan, P. O. Box 41335-1914, Rasht, Iran*

**Abstract**

The density functional theory (DFT) is used to investigate the mechanical properties of pure, fully hydrogenated, semi-hydrogenated, fully fluorinated, and semi-fluorinated germanene sheets, including Young's and bulk moduli, and plastic properties. Also, the electronic properties, namely the planar average electron difference density, are considered to evaluate the bonding characteristics of pure and adsorbed germanenes. The results show that the effect of adsorption decreases the electron accumulation between Ge atoms, i.e. weaker covalent bonds. This weakening in the covalent bonds results in the reduction of Young's and bulk moduli, which are calculated through the second derivation of the total energy versus strain. Furthermore, it is observed that the yield strain of fully fluorinated germanene remains unchanged under uniaxial loading as compared to the pristine structure. Moreover, the yield strain of the germanene is increased under biaxial loading, while it is affected by full hydrogenation adsorption.

## 1. Introduction

Group-IV monolayers, namely graphene, silicene, germanene, and stanene, have gained much attention in the last decade, for their high carrier mobility, strong mechanical characteristics, and structural stability [1–6]. Among these, germanene is an appealing material, which has been synthesized on several substrates, including Pt (111), Au (111), and Al (111), by epitaxial growth [7–9]. Unlike graphene, which is completely flat, the larger atoms of germanene weaken the π−π overlaps, which leads to buckled structures with sp$^2$ and sp$^3$ hybrid orbitals. Hitherto, germanene shares most of the interesting electronic properties of graphene, such as Dirac cone and high carrier mobility [10]. Moreover, it has some advantages over graphene such as stronger spin-orbit coupling, which is important in realization of quantum spin Hall effect in room temperature, and better tenability of the band gaps, which are very vital in valleytronics and optoelectronic applications [11,12].

* Corresponding author,
Email: r_ansari@guilan.ac.ir

There are several methods developed for tuning the physical characteristics of 2D materials on-demand, such as designing heterostructures, applying strain and electric field, vacancy, defection, decoration, etc. [13–22]. For example, Mozvashi, et al. have designed a heterostructure of antimonene and bismuthene in which the optical absorption in the visible region has increased by ~ 4 $m^{-1}$ compared with the mother monolayers. They gained a doubled Young's modulus for the HS, which may bring hope for stronger robustness of nano-scale mirrors [23]. Moreover, Vishkayi and Bagheri show that it is possible to significantly increase the thermoelectrical efficiency of the novel $C_3N$ sheet by applying a strain perpendicular to the electrical transport, which increases potential application of this novel nano-sheet in thermoelectric and optoelectronic devices [24].

Adsorption, especially by hydrogen and halogens is another well-known method for stabilizing and tuning the physical properties of nano-materials [25]. For instance, Mohebpour et al. have suggested that hydrogenation stabilizes the freestanding structure of the newly on-substrate synthesized $Sn_2Bi$ compound [26,27]. Moreover Saffari et al. suggest that chlorination of $CH_3NH_3PbI_3$ perovskite solar cells leads to an increased band gap and optical parameters, such as refractive index and extinction coefficient, which eventually enhances the efficiency of the solar cells [28]. Generally, there are several models of adsorption for 2D materials, including table, chair, boat, and stirrup -like adsorptions, all may also be full and/or semi absorbed. Among these, the chair-like model is famous for the mechanical and optical isotropy, as well as more stability, due to its in-plane symmetry [29–33].

For the germanene, the stability, electronic, and magnetic properties of the full and semi hydrogenated and fluorinated structures have been previously investigated several times [34–37]. For instance, Wang et al. by means of first-principles calculations found transitions in germanene from metallic to magnetic semiconductor and then to nonmagnetic semiconductor, for pristine, semi-hydrogenated, and full-hydrogenated systems, respectively. More importantly, they confirm that these structures are dynamically stable in the ground state based on the phonon mode analysis [35]. Moreover, Trivedi, et al. by DFT calculations investigated the effects of hydrogenation on germanene with different configurations (chair, boat, table, and stirrup) and found that the chair-like structure has the highest binding energy, i.e. are more stable [37].

What's more, Shu et al. by using the DFT, $G_0W_0$ method and Bethe-Salpeter equations investigated the chair-like hydrogenated and fluorinated germanene and found that they are direct band-gap semiconductors, with significant quasi-particle corrections and strong excitonic effects with large exciton binding energies [36]. Also, Liang, et al. investigated the chair-like full and semi hydrogenated, fluorinated and chlorinated germanene, and found that the semi-adsorbed models present magnetic characters, namely, the semi-hydrogenated model is a ferromagnetic semiconductor, while the semi-fluorinated and -chlorinated ones are anti-ferromagnetic semiconductors. Moreover they suggest that the stability of magnetic coupling of these semi-adsorbed germanene models can be modulated by external strain [34].

Despite the mentioned studies about the stability, electronic, and magnetic characteristics, the mechanical properties, such as Young's and bulk moduli, as well as elastic and plastic behavior of these adsorbed germanene structures have not gained sufficient consideration so far. In this paper, we investigated how the bonding characteristics of the germanene are affected by the adsorption, and then we calculated Young's and bulk moduli, as well as the critical strains, which are used to determine when the materials enter the plastic region, through density functional theory (DFT) calculations.

## 2. Computational details

The Spanish code package, SIESTA was employed for all the simulations, which is based on self-consistent density functional theory (DFT) and standard pseudo-potentials [38]. The generalized gradient approximation (GGA) with parameterization of the Perdew-Burke-Ernzerhof (PBE) [39] are applied to evaluate the correlation effects and atomistic geometrical optimizations. The chosen functional (GGA-PBE) has approved to have good compatibility with experiments in geometric and electronic analysis of group IV and V monolayers [40]. The atomic orbital pseudo-potentials of double-ζ plus polarization orbitals (DZP) with an energy shift of 50 meV alongside the split norm of 0.3 were utilized for all of the procedures and the atomic locations were relaxed until the remaining forces on any atom were under 0.015 eVA$^{-1}$. In all the calculations, the reciprocal space was sampled by a mesh of 20×20×1 k points in the Brillouin zone and the mesh cutoff is considered 325 Ry. Moreover, the DFT-D3 correction of Grimme was considered for adsorbed structures to take into account the Van-der Waals interactions [41]. The vacuum height of 17 Å is set for the unit cells to avoid spurious interaction between periodically repeated images. This computational set-ups were chosen by energy convergence of less than $10^{-4}$ eV.

## 3. Results and discussion

### 3.1. Structural configurations

In addition to the pristine germanene, four chair-like absorbed models were considered, including semi and full hydrogenated and fluorinated structures (SH-Ge, FH-Ge, SF-Ge, and FF-Ge, respectively), as explained in Figure 1. Structures were relaxed by the mentioned procedure in the computational details, and the cohesive energies were evaluated through:

$$E_c^{sheet} = \frac{E_{sheet} - 2E_{Ge}}{2}$$

$$E_c^{ads} = \frac{E_{ads} - (2E_{Ge} + mE_X)}{2 + m}$$

(1)

where $E_{sheet}$ is the total energy of the pure germanene, $E_{ads}$ is the total energy of the adsorbed germanene, $E_{Ge}$ is the total energy of the isolated germanium atom, regarding the spin polarization, $E_X$ is the total energy of the ad-atom (hydrogen or fluorine), regarding the spin polarization, and m is the number of ad-atoms in the full or semi adsorbed structures (2 or 1). According to Eq. (1), the more negative cohesive energies stand for more stability of the structures.

The structural parameters, including the lattice constants, Ge-Ge bond lengths, buckling, adsorption heights, and the cohesive energies are presented in Table 1, which are in good agreement with the previous works on the pure and absorbed Ge structures. According to our calculated cohesive energies, and previous studies analyses including phonon dispersion [35,37], it is deducible that all the considered structures are thermodynamically stable. Now we turn our attention into the electronic properties of these structures to gain information about the bonding characteristics.

### 3.2. Bonding characteristics

The atoms are bound together by electrons, therefore the electronic properties are the keys to know the mechanism of the atomic bonds. The average electron difference density ($\langle \delta n(r) \rangle$), is significant to understand the bonding nature of the materials [42,43]. We calculated the average electron difference density of the pure and adsorbed germanene to have insight into the bonding characteristics of these structures.

Figure 2 shows the average electron difference density, $\langle \delta n(r) \rangle$ for the pure and adsorbed germanene structures. As it is clear, pure germanene has a very intensive accumulation of electrons between Ge atoms, which is signature of strong covalent bonds. However, by adsorption of H and F atoms, no matter full or semi, the ad-atoms attract a significant share of electrons and the accumulation within the Ge atoms decreases. This is due to the larger electronegativity of the F (3.98) and H (2.20) in comparison with Ge (2.01) [44]. Hence, the strength of covalent bonds among Ge atoms are reduced due to the less participation of electrons in nanosheets. It is predicted that the adsorbed germanene should have lower Young's modulus than that of the pure one, which makes them softer materials.

To more deeply see the mechanism of how the atoms are connected together, we performed a partial density of states (PDOS) analysis. Figure 3 shows the total and orbital projected PDOS for the pure, semi, and full adsorbed germanenes. As it is clear, the pure and semi-adsorbed germanenes are metallic, where the full-adsorbed germanenes are semiconductors. Moreover, we see a mild Dirac cone at the Fermi level for pure germanene. All of these electronic characteristics are compatible with previous works [34–37,45].

In aspects of bonding characteristics, which is bound to the mechanical properties, orbital hybridization plays an important role, which can be deduced from the PDOS [20,46]. As can be seen in pure germanene, the p orbitals are dominant, where the s orbitals do not have significant contribution. The two Ge atoms have similar contributions, which shows strong in-plane $\sigma$ hybridization between the p orbitals of the two Ge atoms. In the case of semi-adsorbed models, the symmetry of Ge-p orbitals in the xy plane breaks and the two Ge-p orbitals make different contributions. In other words, the adsorbed Ge atom (Ge$^*$) has less contribution in the DOS than the non-absorbed one (Ge). This is due to additional bonds between Ge$^*$ and the adatom (H or F), which should lead to less in-plane hybridization and weaker atomic bonds.

Moreover, in the case of full-adsorption, both Ge atoms have an additional bond with the adatoms, therefore the symmetry of p orbitals in the xy plane reserves and the contribution of the two Ge atoms imitate each other. In the FH-Ge model, the H-s orbital does not play significant role, therefore the bonding characteristics of the FH-Ge models should be similar to the pure-Ge. Although in the FF-Ge model, contribution of out-of-plane F-p orbital is significant, which is predicted to moderately lower the strength of the Ge-Ge in-plane bonds.

To extract more detailed data, the mechanical characteristics of adsorbed germanene, in the following sections we calculate elastic and plastic properties of the pure and adsorbed nano-sheets.

*3.3. In-plane Young's modulus*

The in-plane Young's (elastic) modulus of germanene structures are calculated in this part and the results are compared. The elastic modulus of the pure and adsorbed germanene is defined as $Y_s = E * t$, where $E$ is Young's modulus and $t$ is the thickness. The advantage of $Y_s$ over E is that it ignores the impact of thickness, which can bring more converged results for different thicknesses chosen in similar studies. Eventually, based on the second derivation of the total energy ($E_s$) versus strain ($\varepsilon$), the in-plane elastic modulus can be calculated through [47,48]:

$$Y_s = \frac{1}{A_0} \frac{\partial^2 E_s}{\partial \varepsilon^2} \qquad (2)$$

$A_0$ demonstrates the unstrained state area of the unit cell. For this purpose, structures are subjected to the uniaxial tensile and compressive loading within the range of -5% to 5% with paces of 1% for longitudinal and transverse directions, and the energy values are recorded. The energies are the plotted versus to the applied strain, which forms sagittal curves for each individual structure and were used for calculating the $\frac{\partial^2 E_s}{\partial \varepsilon^2}$. These curves are presented in Figure 4 and 5 for the two directions, alongside with their second derivatives. Moreover, the calculated Young's moduli of the pure and adsorbed nanosheets, as well as the percentage of the difference with respect to the pure structure are presented in Table 2.

According to these results, it can be said that the adsorption decreases Young's moduli of germanene in both longitudinal and transverse directions. The largest reduction in the longitudinal and transverse directions is observed for the SF-Ge, while the lowest reduction occurs for FH-Ge structure in both directions. Moreover, the difference between the two directions is negligible, hence it can be concluded that the pure and the adsorbed structures are isotopic and adsorption would not affect this behavior. These results supports our analytic explanation about the electron density and partial DOS in section 3.2.

*3.4. Bulk modulus*

In this section, the biaxial loading is applied along both directions in order to evaluate the bulk moduli of pristine and adsorbed structures. To calculate the bulk moduli of structures, both directions are subjected to compressive and tensile loadings. The bulk modulus can be computed using the presented relation [47,48]:

$$B = A_0 \frac{\partial^2 E_s}{\partial A^2} \tag{3}$$

Where $A_0$ and $A$ are the unstrained and strained area of the nanosheets, respectively. Following the same steps used to evaluate the elastic modulus, the related curves of the germanene structures are obtained and presented in Figure 6. Using these curves, $\frac{\partial^2 E_s}{\partial A^2}$ were calculated and substituted in Eq. (2) to acquire the bulk moduli of the structures. The calculated bulk moduli, for the pure and the adsorbed structures are given in Table 3.

According to these results, it is clear that in all cases the adsorption cause a decrease in the bulk moduli. Moreover, the greatest and the lowest reductions are observed for SF-Ge (-47.50%) and FF-Ge (-29.14%) respectively.

As mentioned in section 3.2, because of the lower electronegativity of Ge, compared to H and F, the distribution of electrons reduces in adsorbed germanene, and the electron accumulation decreases among Ge atoms and increases around the ad-atoms. This reduces the bonding strength of Ge-Ge as well as the elastic and bulk moduli. Moreover, the bonding strength are even weaker in the fully fluorinated germanene for the higher electronegativity of the F atoms compared with H, therefore they possess lower elastic and bulk moduli. The semi adsorbed structures have less planar symmetry, therefore less electrons participate in the xy plane, which results to weaker covalent bonds, compared with the full adsorbed structures. This is the key point in more reduction in the elastic and bulk moduli of the semi adsorbed structures. The reduction in the elastic and bulk moduli of structures caused by adsorption are analogous with previously works on silicene [43,49,50], bismuthene [47], arsenene [22,48], and antimonene [51], adsorbed systems which proves the validity of the present study.

*3.5. Plastic properties*

To evaluate plastic properties of nanosheets, loading is further increased which eventually cause three different sections to appear. The three sections are separated by two different strains, i.e. first,$\varepsilon_{c1}$ and second,$\varepsilon_{c2}$ critical strains. The first one is where $dE_T(\varepsilon)/d\varepsilon$ has its maximum value and the region before it, demonstrated the harmonic region. The second one ($\varepsilon_{c2}$), is where the strain energy would be constant with no change in its value or it reaches its highest value.

Furthermore, the region between $\varepsilon_{c1}$ and $\varepsilon_{c2}$ can be defined as the inharmonic region. The whole area from the point when the strain is zero until the energy reaches the second critical strain is defined as the elastic region. Finally, for the strains larger than the second critical strain, the nanosheet enters the plastic region. The studied critical strains in the current study are shown in Figure 7 and 8, for all nanosheets under the uniaxial, and biaxial loadings, respectively. The uniaxial loading along the transverse direction was ignored, for the isotropy of the structures, which was said before in section 3.3. In addition, the values of the critical strains acquired from the plots, are presented in Table 4.

Under uniaxial loadings, the results show that the width of the harmonic region($\boldsymbol{\varepsilon c_1}$), reduces in SH-Ge and increases in FF-Ge structure, while it remains unchanged in other cases. In addition, except for FF-Ge structure, adsorption decreases the yield strain($\boldsymbol{\varepsilon c_2}$). Moreover, the inharmonic region ($\varepsilon c_2 - \varepsilon c_1$) is reduced in all of the adsorbed structures with the highest reduction in FF-Ge and the lowest reduction in FH-Ge structures.

For the biaxial loadings, adsorption leads to a reduction in the first critical strain in all structures. However, the second critical strain of the structure increases for the FH-Ge, while other adsorptions decrease the second critical strain of the nanosheet. The highest and lowest reductions are observed for SH-Ge and FF-Ge nanosheets, respectively. Surprisingly, fully hydrogenation also increases the inharmonic region of the structure while in other adsorptions the width of this area is decreased.

## 4. Conclusion

By means of density function theory, we firstly investigated the bonding characteristics of the pure and adsorbed germanene. The average electron difference density shows that the adsorption leads to a decrease in covalent bonds of the germanene. This decrease can be due to the reduction of electron distribution in the xy plane, for the high electronegativity of the ad-atoms. The changes are also approved by PDOS, which shows alteration of atomic orbitals hybridization. Moreover, these results are covered by the calculation of the elastic and bulk moduli, which are also reduced due to the adsorption. While both elastic and bulk moduli of the adsorbed structures are decreased, similar result is not observed for the plastic behavior of the adsorbed structures. The yield strain of the FF-Ge remained unchanged under uniaxial loading in comparison to the pure structure. In addition, the yield strain of the germanene increases while affected by fully hydrogenation adsorption under the biaxial loading. Overall, the results obtained in the current study is valuable for the future use in mechanical sensors and nanoscale systems.

## Table Captions

**Table 1:** Structural parameters of the pure and adsorbed germanene, including lattice constant (a), Ge-Ge bond length (d), buckling (Δ), adsorption height (h), and cohesive energy ($E_c$), in comparison with previous studies. These parameters are also explained in Figure 1.

**Table 2:** Young's moduli along with the longitudinal and transverse directions of the pure and adsorbed germanene structures. The percentage of the difference with the pure structure is also shown.

**Table 3:** Bulk moduli of the pure and adsorbed germanene structures. The percentage of the difference with the pure structure is also shown.

**Table 4:** $\varepsilon c1$ and $\varepsilon c2$ of the pristine and the adsorbed germanene under uniaxial and biaxial strains.

# Figure Captions

**Figure 1:** Structural configurations of the pure and adsorbed germanene. The structural parameters, the longitudinal and transverse directions are also explained.

**Figure 2:** (Color) The average electron difference density for the pure and adsorbed germanene. The electron accumulation and depletion are shown in red and blue colors, respectively.

**Figure 3:** Total and orbital projected density of states (DOS) calculated for pure, semi, and full adsorbed germanenes.

**Figure 4:** Curves of the strain energy difference vs. strain, along the armchair direction for the pure and the adsorbed Ge structures. The energy of the unstrained structures are shifted to zero.

**Figure 5:** Curves of the strain energy difference vs. strain, along with the zigzag direction for the pure and the adsorbed Ge structures. The energy of the unstrained structures are shifted to zero.

**Figure 6:** Curves of the strain energy difference vs. surface area for the pure and the adsorbed Ge structures. The energy of the unstrained structures are shifted to zero.

**Figure 7:** The first and the second critical strains of the pristine and adsorbed germanene under uniaxial strains along with the armchair direction. The energy of the unstrained structures are shifted to zero.

**Figure 8:** The first and the second critical strains of the pristine and adsorbed germanene under biaxial strains. The energy of the unstrained structures are shifted to zero.

# Tables

**Table 1**: Structural parameters of the pure and adsorbed germanene, including lattice constant (a), Ge-Ge bond length (d), buckling (Δ), adsorption height (h), and cohesive energy ($E_c$), in comparison with previous studies. These parameters are also explained in Figure 1.

| Nano-sheet | a (Å) | d (Å) | h (Å) | Δ (Å) | $E_c$ (eV/atom) | Reference |
|---|---|---|---|---|---|---|
| Pure Ge | 4.09 | 2.46 | --- | 0.69 | -3.70 | This work |
|  | 4.03 | 2.44 | --- | 0.74 |  | Ref. [37] |
| FH-Ge | 4.12 | 2.49 | 1.57 | 0.75 | -3.17 | This work |
|  | 3.91 | 2.40 |  | 0.82 |  | Ref. [37] |
|  | 4.02 | 2.44 | 1.52 | 0.75 | -3.70 | Ref. [52] |
| SH-Ge | 4.23 | 2.53 | 1.59 | 0.81 | -3.03 | This work |
|  | 4.12 | 2.51 | 1.58 | 0.77 |  | Ref. [36] |
|  | 4.11 | 2.49 | 1.58 | 0.75 |  | Ref. [53] |
| SF-Ge | 4.20 | 2.54 | 1.83 | 0.75 | -4.00 | This work |
|  | 4.16 | 2.51 | 1.79 | 0.73 |  | Ref. [54] |
| FF-Ge | 4.29 | 2.55 | 1.81 | 0.62 | -3.89 | This work |
|  | 4.22 | 2.49 | 1.78 | 0.61 | -4.33 | Ref. [52] |
|  | 4.10 | 2.40 | 1.7 | 0.61 |  | Ref. [36] |

**Table 2**: Young's moduli along with the longitudinal and transverse directions of the pure and adsorbed germanene structures. The percentage of the difference with the pure structure is also shown.

| Structure | Longitudinal Young's modulus (N/m) | Difference (%) | Transverse Young's modulus (N/m) | Difference (%) |
|---|---|---|---|---|
| Ge | 47.87 | --- | 47.86 | --- |
| SH-Ge | 43.63 | -8.85 | 43.52 | -9.07 |
| FH-Ge | 46.42 | -3.02 | 46.73 | -2.36 |
| SF-Ge | 30.53 | -36.22 | 30.32 | -36.64 |
| FF-Ge | 36.15 | -24.48 | 36.13 | -24.50 |

**Table 3**: Bulk moduli of the pure and adsorbed germanene structures. The percentage of the difference with the pure structure is also shown.

| Structure | Bulk modulus (N/m) | Difference (%) |
|---|---|---|
| Ge | 26.04 | --- |
| SH-Ge | 14.85 | -42.97 |
| FH-Ge | 18.46 | -29.10 |
| SF-Ge | 13.67 | -47.50 |
| FF-Ge | 18.45 | -29.14 |

**Table 4**: $\varepsilon_{c1}$ and $\varepsilon_{c2}$ of the pristine and the adsorbed germanene under uniaxial and biaxial strains.

| Structure | Uniaxial strain | | Biaxial strain | |
|---|---|---|---|---|
| | $\varepsilon_{c1}$ | $\varepsilon_{c2}$ | $\varepsilon_{c1}$ | $\varepsilon_{c2}$ |
| Ge | 0.21 | 0.48 | 0.24 | 0.45 |
| FH-Ge | 0.21 | 0.45 | 0.21 | 0.51 |
| SH-Ge | 0.15 | 0.24 | 0.09 | 0.12 |
| SF-Ge | 0.21 | 0.27 | 0.18 | 0.21 |
| FF-Ge | 0.24 | 0.48 | 0.18 | 0.36 |

# Figures

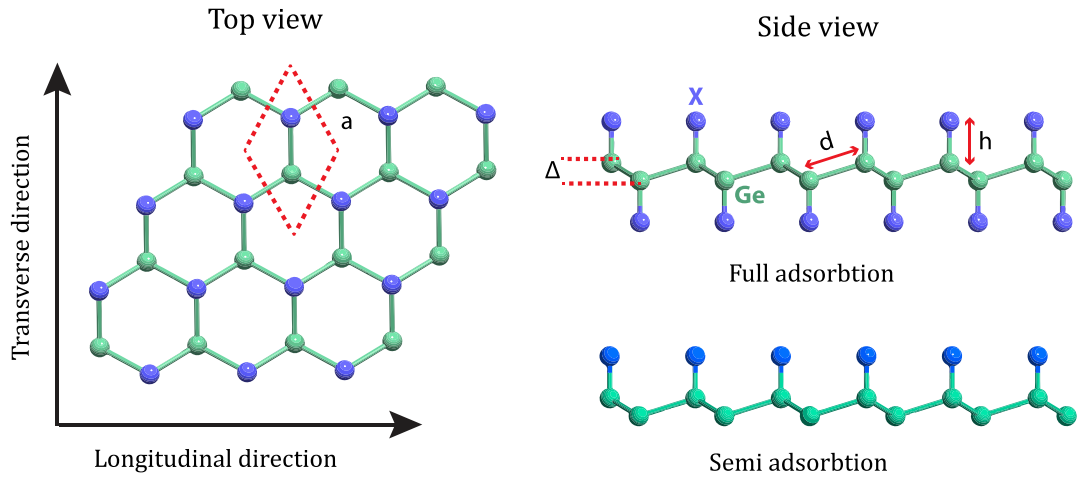

**Figure 1**: Structural configurations of the pure and adsorbed germanene. The structural parameters, the longitudinal and transverse directions are also explained.

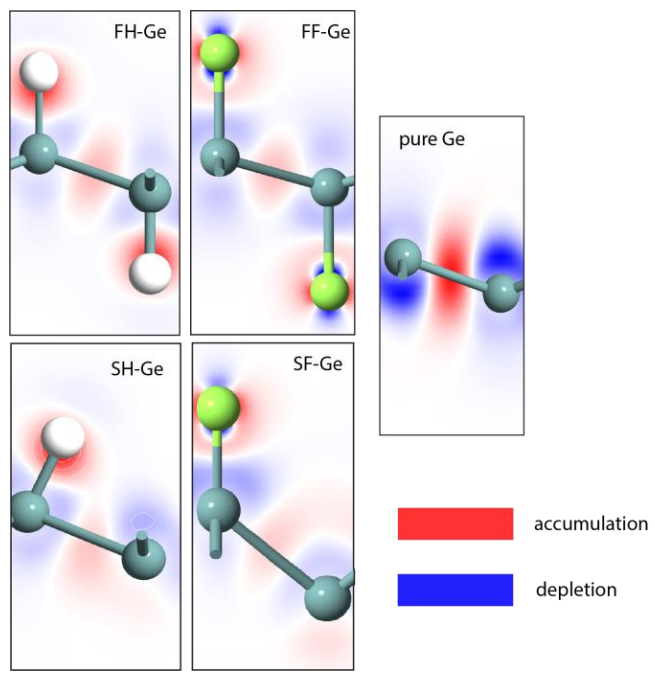

**Figure 2**: (Color online) The average electron difference density for the pure and adsorbed germanene. The electron accumulation and depletion are shown in red and blue colors, respectively.

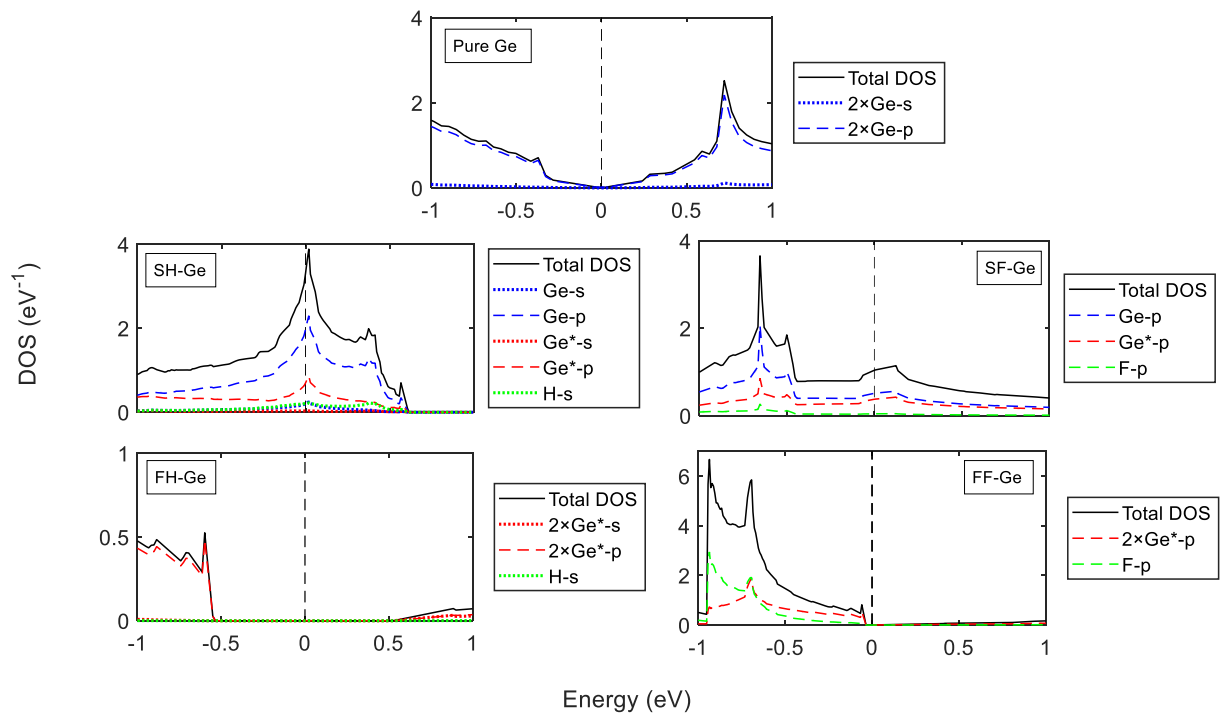

**Figure 3:** (Color online) Total and orbital projected density of states (DOS) calculated for pure, semi, and full adsorbed germanenes. The adsorbed Ge atoms are clarified by stars (*).

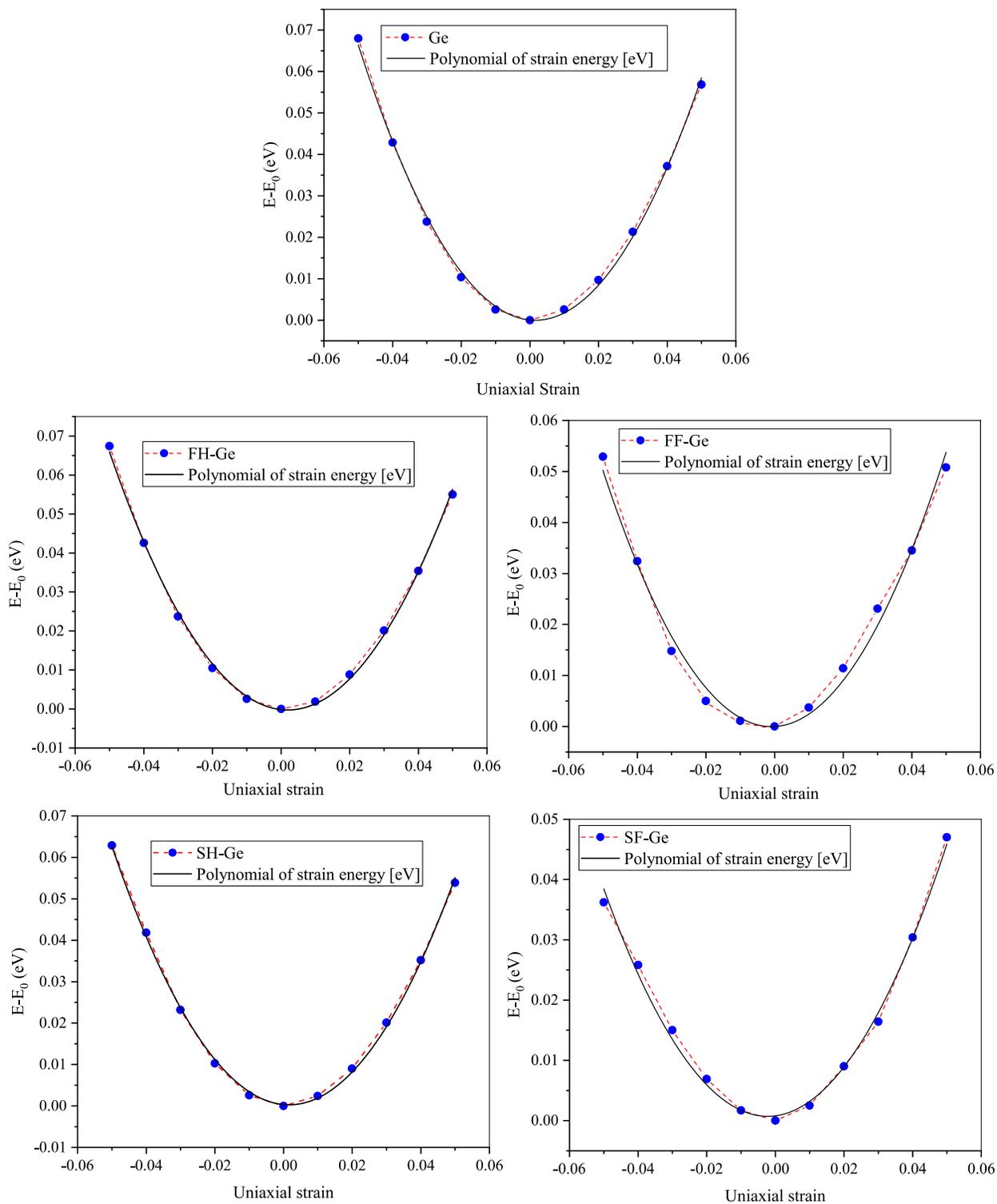

**Figure 4**: Curves of the strain energy difference vs. strain, along the armchair direction for the pure and the adsorbed Ge structures. The energy of the unstrained structures are shifted to zero.

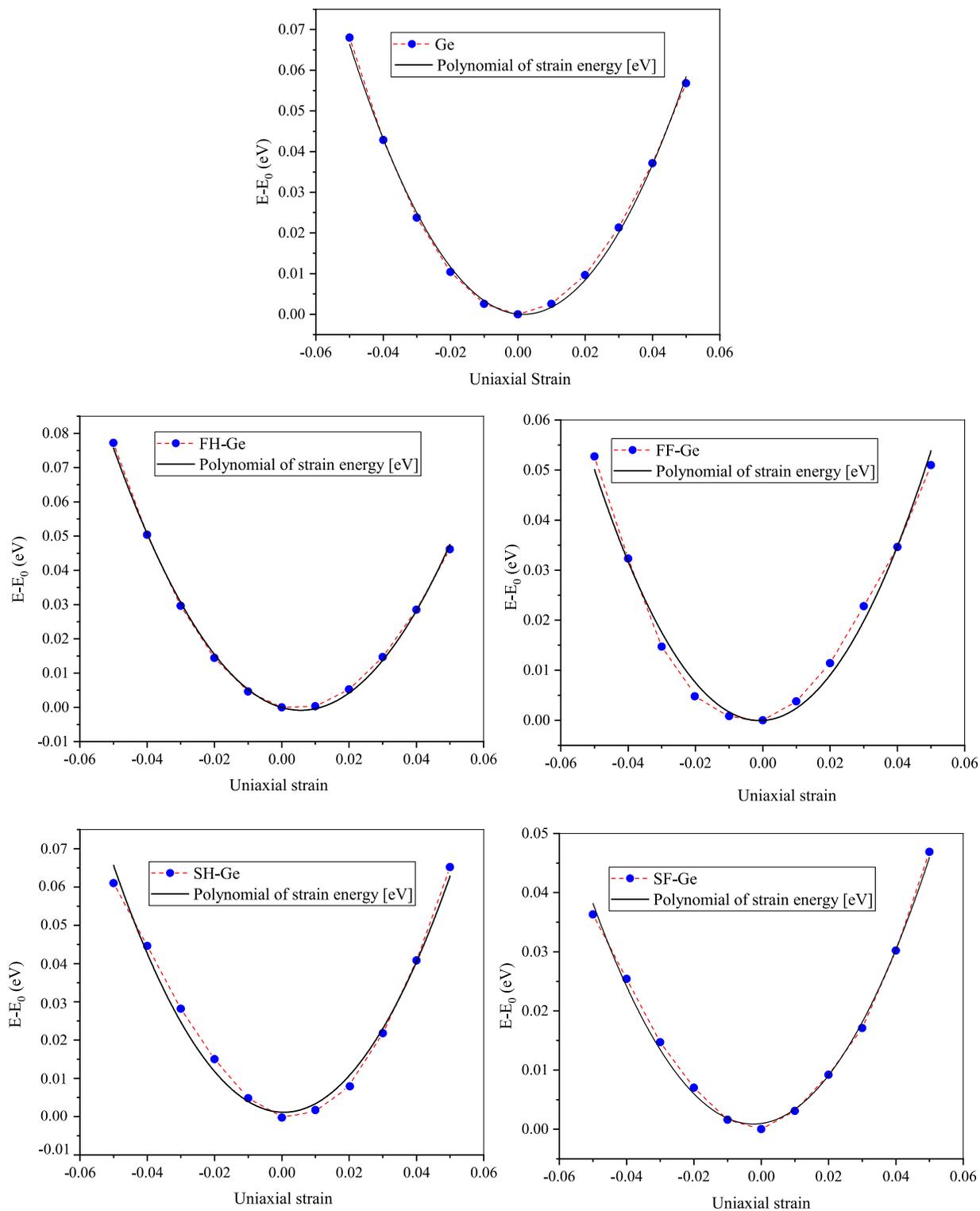

**Figure 5**: Curves of the strain energy difference vs. strain, along with the zigzag direction for the pure and the adsorbed Ge structures. The energy of the unstrained structures are shifted to zero.

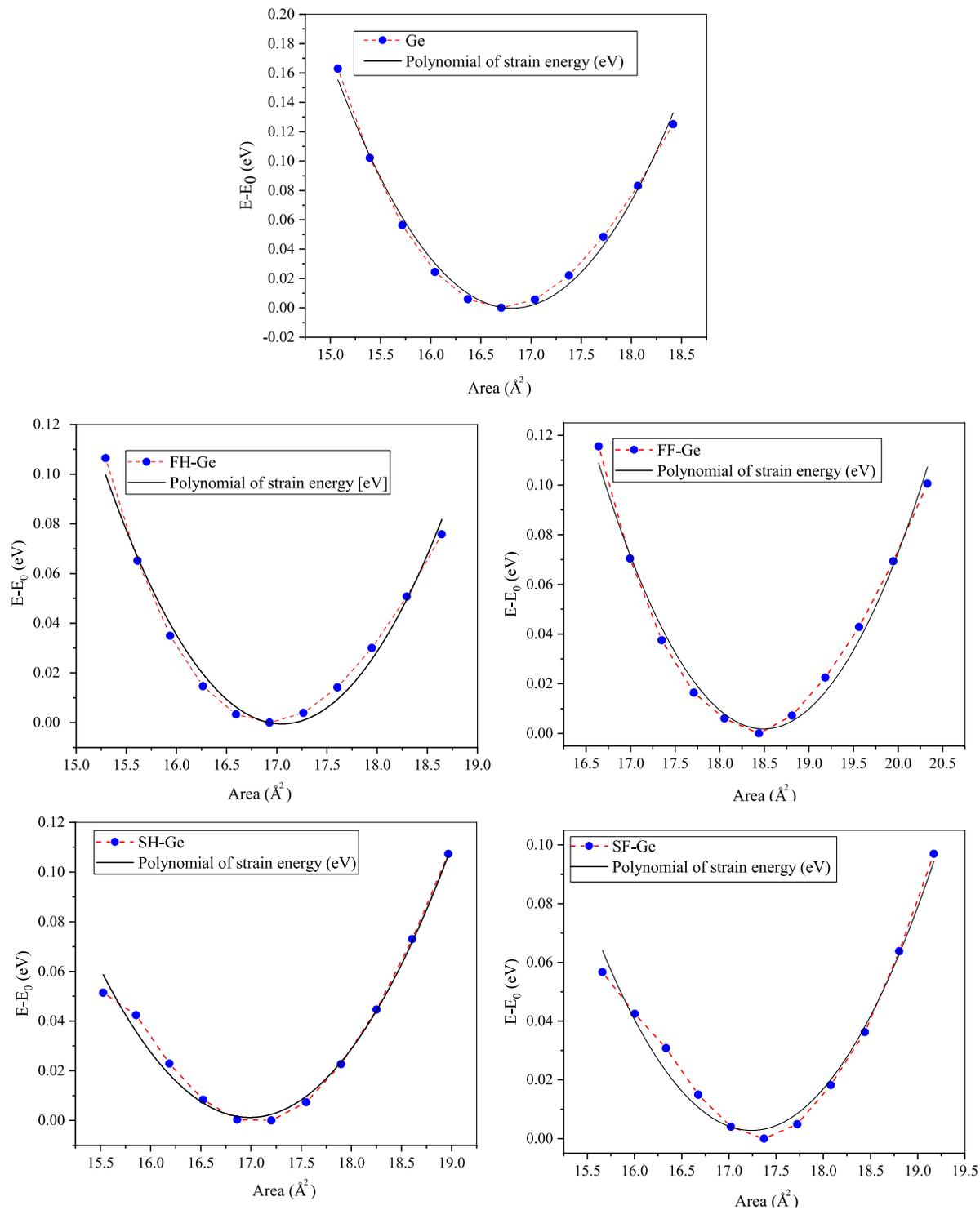

**Figure 6**: Curves of the strain energy difference vs. surface area for the pure and the adsorbed Ge structures. The energy of the unstrained structures are shifted to zero.

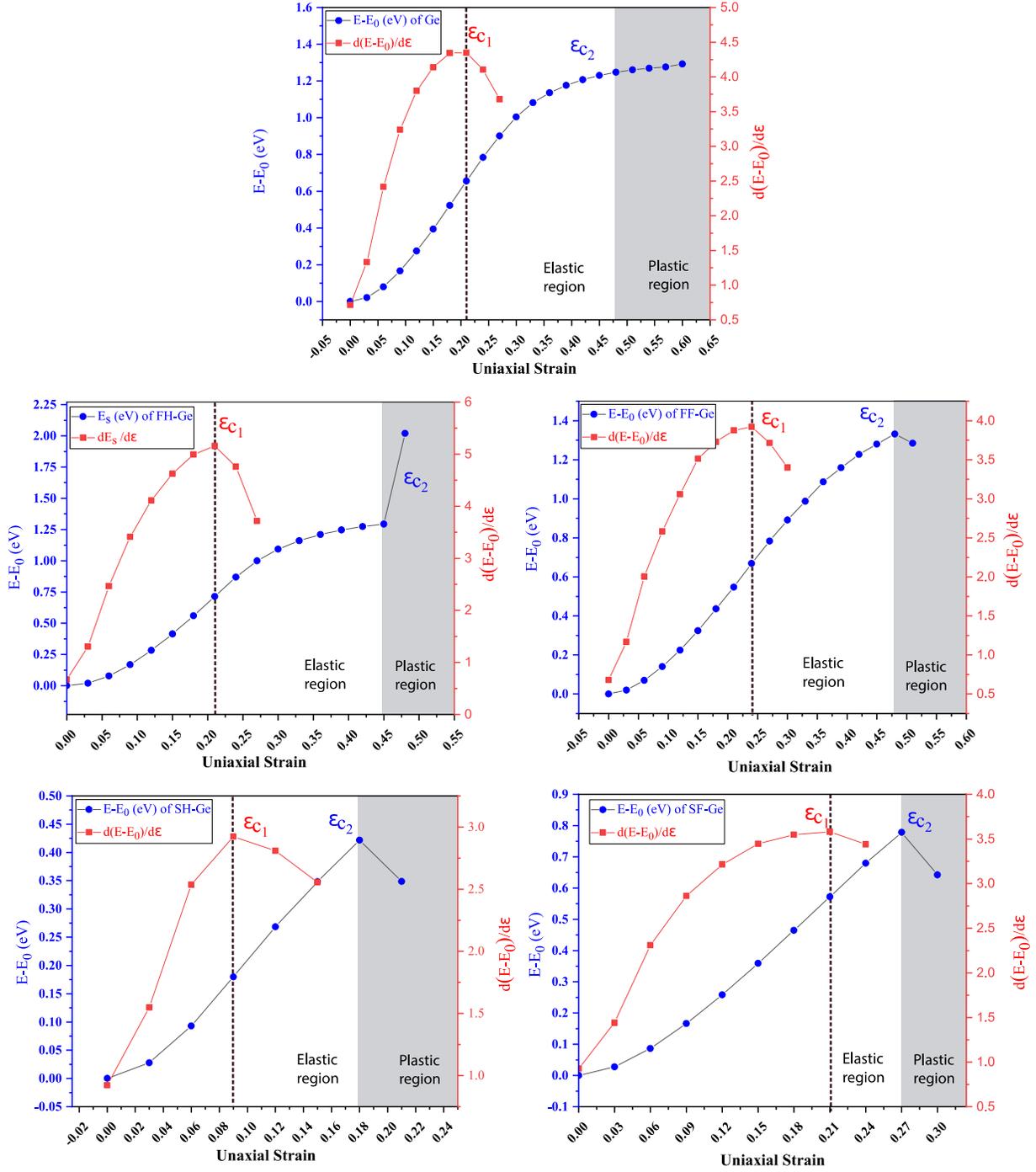

**Figure 7**: The first and the second critical strains of the pristine and adsorbed germanene under uniaxial strains along with the armchair direction. The energy of the unstrained structures are shifted to zero.

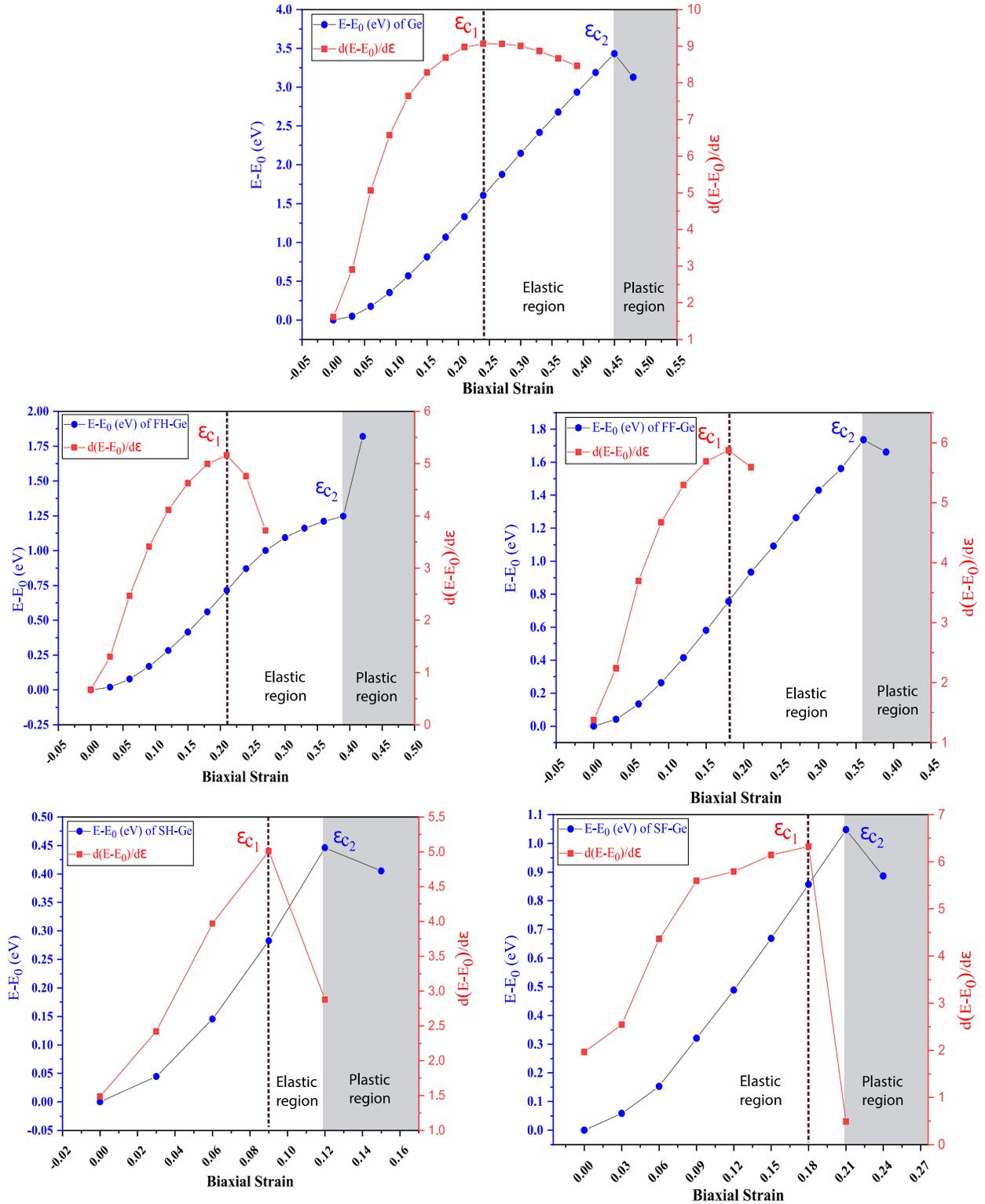

**Figure 8**: The first and the second critical strains of the pristine and adsorbed germanene under biaxial strains. The energy of the unstrained structures are shifted to zero.